\documentclass[journal=jpcafh,manuscript=article,layout=traditional]{achemso}

\usepackage[version=3]{mhchem} 
\usepackage[T1]{fontenc}       

\usepackage{lipsum}
\usepackage{color}
\usepackage{gensymb}

\newcommand{\RNum}[1]{\uppercase\expandafter{\romannumeral #1\relax}}

\graphicspath{{./images/}}

\author{Markus Schneider}
\affiliation[FHI Berlin]{Fritz-Haber-Institut der Max-Planck-Gesellschaft, Theory Department, Faradayweg 4-6, D-14195 Berlin, Germany}
\altaffiliation{Both authors contributed equally.}
\author{Chiara Masellis}
\affiliation[EPF Lausanne]{Ecole Polytechnique F\'ed\'erale de Lausanne, Laboratoire de Chimie Physique Mol\'eculaire, EPFL SB ISIC LCPM, Station 6, CH-1015 Lausanne, Switzerland}
\altaffiliation{Both authors contributed equally.}
\author{Thomas Rizzo}
\email{thomas.rizzo@epfl.ch}
\affiliation[EPF Lausanne]{Ecole Polytechnique F\'ed\'erale de Lausanne, Laboratoire de Chimie Physique Mol\'eculaire, EPFL SB ISIC LCPM, Station 6, CH-1015 Lausanne, Switzerland}
\author{Carsten Baldauf}
\email{baldauf@fhi-berlin.mpg.de}
\affiliation[FHI Berlin]{Fritz-Haber-Institut der Max-Planck-Gesellschaft, Theory Department, Faradayweg 4-6, D-14195 Berlin, Germany}

\title{Kinetically Trapped Liquid-State Conformers of a Sodiated Model Peptide Observed in the Gas Phase}

\begin{document}


\begin{abstract}
\noindent
We investigate the peptide \ce{AcPheAla_5LysH+}, a model system for studying helix formation in the gas phase, in order to fully understand the forces that stabilize the helical structure.
In particular, we address the question of whether the local fixation of the positive charge at the peptide's C-terminus is a prerequisite for forming helices by replacing the protonated C-terminal Lys residue by Ala and a sodium cation.
The combination of gas-phase vibrational spectroscopy of cryogenically cooled ions with molecular simulations based on density-functional theory (DFT) allows for detailed structure elucidation.
For sodiated \ce{AcPheAla_6}, we find globular rather than helical structures, as the mobile positive charge strongly interacts with the peptide backbone and disrupts secondary structure formation.
Interestingly, the global minimum structure from simulation is not present in the experiment. 
We interpret that this is due to high barriers involved in re-arranging the peptide-cation interaction that ultimately result in kinetically trapped structures being observed in the experiment.
\end{abstract}

\vspace{1cm}

\begin{center}
\fbox{\includegraphics[height=5cm]{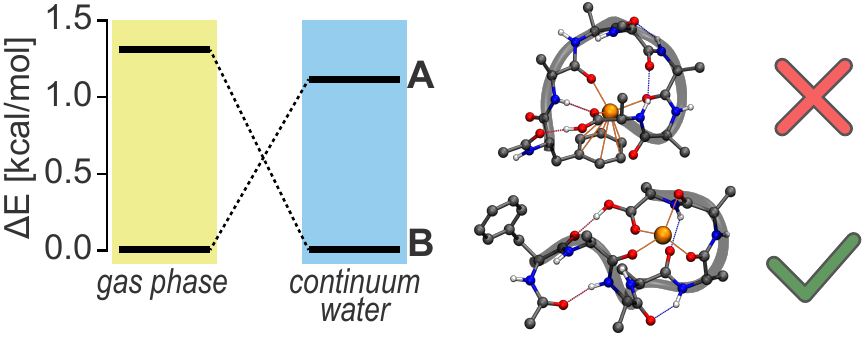}}
\end{center}


\newpage
\section{Introduction}
\noindent
Helical secondary structural motifs, such as $\mathrm{\alpha}$ and $\mathrm{3_{10}}$, are common in proteins.\cite{Barlow1988}
In solution, helix propensity is determined both by intramolecular interactions and protein-solvent interaction. 
Gas-phase systems offer the opportunity to study the ``undamped'' intramolecular interactions that shape peptides, thereby shedding light on intrinsic helix propensities and bonding interactions. 
Gas-phase helices have been investigated using ion mobility spectrometry~\cite{Jarrold1998,Jarrold1999,Jarrold2004} and vibrational spectroscopy~\cite{Mons2005,Rizzo2007,Rizzo2008,Rizzo2009,Rizzo2015,Scheffler2010,Scheffler2015a,Scheffler2015b}. 
The combination of these experimental techniques with molecular simulations based on density-functional theory (DFT) allows for structure elucidation, as it helps to interpret experimentally obtained spectra.
Moreover, a rigorous experiment-theory comparison allows for the assessment of the accuracy and predictive power of simulation approaches.\cite{Rossi2015}.

Pioneering ion-mobility experiments in the group of Jarrold\cite{Jarrold1998,Jarrold1999} examined the role of N- and C-terminal residues on gas-phase helix formation for the sequences \ce{Ala_nH+}, \ce{AcLysAla_nH+}, and \ce{AcAla_nLysH+}. 
They concluded that \ce{Ala_nH+} and \ce{AcLysAla_nH+} adopt globular conformations in the gas phase independent of the length of the amino-acid chain while \ce{AcAla_nLysH+} is helical for $\mathrm{n}>8$.\cite{Rossi2013}
The identities of these structures were confirmed by theoretical and experimental vibrational spectroscopy in the work of Rossi \textit{et al.}~\cite{Scheffler2010} and Schubert \textit{et al.}~\cite{Scheffler2015b}, respectively. 
Similar studies focused on peptides of the form \ce{AcPheAla_nLysH+} with $\mathrm{n}=1\text{--}5,10$, where phenylalanine provides a UV chromophore, which allows for conformer-specific IR-UV double resonance spectroscopy.\cite{Rizzo2007,Rizzo2008,Rizzo2009,Rizzo2015}
Vibrational signatures of individual conformers add a new dimension to peptide structural analysis beyond the orientationally averaged collisional cross section provided by ion mobility.
In these experiments, the number of residues necessary to form a helix was found to be six~\cite{Rizzo2009,Rossi2013}, but much of the hydrogen bonding pattern responsible for the formation of this motif is already present even with only three residues~\cite{HLCA:HLCA201200436,Rizzo2015}.  
In conjunction with computational vibrational spectroscopy based on DFT,\cite{Rizzo2009,Rizzo2015,Scheffler2010,Rossi2014} such spectra allowed for determining detailed molecular structures and critically examining evidence for helix formation of peptides in isolation.

\begin{figure}
\includegraphics[width=1.\textwidth]{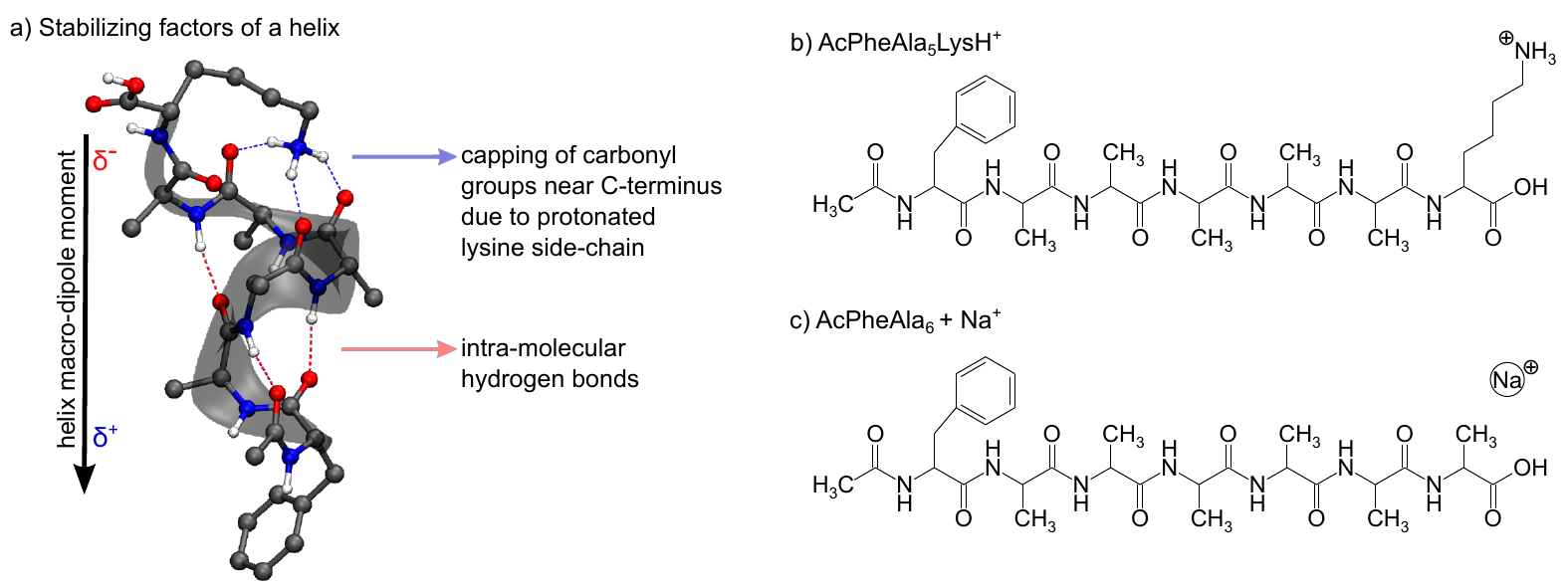}%
\caption{\label{Figure01} Illustration of helix-stabilizing factors for peptides in the gas phase (a) and structural formulas of (b)~\ce{AcPheAla5LysH+} and (c)~\ce{AcPheAla6 + Na+}.}
\end{figure}


Figure~\ref{Figure01}(a) illustrates the helix-stabilizing factors in polyalanine peptides, shown for the specific case of \ce{AcPheAla5LysH+}.
Work by the groups of Jarrold~\cite{Jarrold1998,Jarrold1999}, Rizzo,~\cite{Rizzo2007,Rizzo2008,Rizzo2009,Rizzo2015} and Blum~\cite{Scheffler2015b} showed that intramolecular hydrogen bonds play an important role and that the design concept can even be transferred to non-natural peptides.\cite{Scheffler2015a}
Hoffmann \textit{et al.}~\cite{Hoffmann2016} could show that deleting a single hydrogen bond had little impact on the overall helix stability.
In addition to their energetic stability, hydrogen bonds are aligned in helices, and the resulting macro-dipole favorably interacts with the positive charge of the protonated lysine (\ce{Lys}) side-chain at the C-terminus.
Moreover, the capping of the ``dangling'' carbonyl groups near the C-terminus by the \ce{Lys} side-chain provides additional stability.

To investigate the importance of the charge fixed at the C-terminus, we focus on the well-studied system~\cite{Rizzo2009,Rossi2014} of \ce{AcPheAla5LysH+} and compare it to \ce{AcPheAla6 + Na+} (Figures~\ref{Figure01}(b) and~\ref{Figure01}(c), respectively).
In the latter, \ce{Lys} is formally replaced by alanine (\ce{Ala}) and a sodium cation (\ce{Na+}) in order to introduce a freely movable positive charge.
The resulting rich possibilities for electrostatic interaction can locally disrupt hydrogen-bonding networks and induce unconventional backbone conformations.\cite{Baldauf2013,Ropo2016a,Ropo2016b,DeSandip2017}
Consequently, the cation-binding site, and hence the conformation as a whole, is not \textit{a priori} obvious. 
Ion mobility studies on metallated peptides (\textit{e.g.} sodiated species of \ce{Ala_n + M+}~\cite{Jarrold2000}) suggest that the cation plays the same role as the charged Lys side-chain in \ce{AcAla_nLysH+} for peptides with $\mathrm{n}>12$. 
For shorter peptides, calculated collisional cross sections (CCS) for globular and helical structures are both in agreement with the experimental CCS, preventing a definitive structural assignment.
In the present work, we couple IR-UV double resonance spectroscopy and theory in order to unravel the structure of the system of \ce{AcPheAla6 + Na+} with the aim of understanding whether a freely movable cation is sufficient to stabilize helix formation or if the C-terminal localization is a prerequisite for that.


\section{Experimental Setup}

The experimental setup has been described in detail elsewhere~\cite{Rizzo2010}. In brief, a nano-electrospray ion source is combined with a cooled ion trap ($4\,\mathrm{K}$) for spectroscopic studies of gas-phase ions. 
Conformer-selective IR spectra are recorded by applying IR-UV double resonance.
A measurement is performed by fixing the wavenumber of the UV laser to a line in the electronic spectrum and scanning the wavenumber of an infrared laser.
When the IR pulse is in resonance with a vibrational transition of the ion, part of the population is removed from the ground state, leading to a decrease in UV-induced fragmentation. 
Scanning the IR wavenumber, one obtains a conformer-specific vibrational spectrum. 
Performing the same experiment on each line of the electronic spectrum allows for assignment of each UV spectral feature to a particular conformer.


\section{Computational Methods}

The applied conformational search algorithm is similar to the one used by Rossi \textit{et al.}\cite{Rossi2014} 
First, a global conformational search is performed on the force field (FF) level using \texttt{CHARMM22}~\cite{CHARMM22} and \texttt{OPLS-AA}~\cite{OPLS1,OPLS2}, separately.
To that end, a basin-hopping approach~\cite{BasinHopping} was applied using the \texttt{scan} program of the \texttt{TINKER} molecular modeling package~\cite{TINKER1,TINKER2}.
For the system of \ce{AcPheAla5LysH+} (\ce{AcPheAla6 + Na+}) $603\,280$ ($626\,829$) conformers were found using \texttt{CHARMM22} and $643\,938$ ($635\,120$) conformers were found using \texttt{OPLS-AA}.
Single-point energy calculations at the generalized-gradient approximated (GGA) DFT level of theory have been performed for \textit{all} these FF conformers.
All DFT calculations were done using the all-electron/full-potential electronic structure code package \texttt{FHI-aims}~\cite{FHIaims,FHIaimsRI,FHIaimsELPA}.
To be more precise, energies were computed at the PBE+vdW level, \textit{i.e.} using the PBE~\cite{PBE} functional and a pair-wise van der Waals correction scheme (vdW)~\cite{TS}.
Furthermore, \texttt{FHI-aims}-specific \texttt{tier 1} basis sets and \texttt{light} settings have been used that are provided out-of-the-box to control the computational accuracy intended to give reliable energies energy for screening purposes~\cite{FHIaims}.
For the two FFs individually, the 500 conformers with the lowest FF energy and the 500 conformers with the lowest DFT energy, \textit{i.e.} a grand total of 2000 conformers, have been selected.
The 2000 selected conformers were then geometry optimized at the PBE+vdW level using \texttt{tier 1} basis sets and \texttt{light} settings.
A hierarchical clustering scheme was applied in order to rule out duplicates.
Further relaxation was then accomplished at the PBE+vdW level using \texttt{FHI-aims}-specific \texttt{tier 2} basis sets and \texttt{tight} settings that are intended to provide $\mathrm{meV}$-level accurate energy differences~\cite{FHIaims}, \textit{i.e.} within $0.02\,\mathrm{kcal/mol}$.
After clustering, this resulted in 324 (159) conformers for the system of \ce{AcPheAla5LysH+} (\ce{AcPheAla6 + Na+}) in the low-energy region, \textit{i.e.} within $6\,\mathrm{kcal/mol}$ from the global minimum. 
These conformers were then again locally refined at the PBE0+MBD level, \textit{i.e.} using the hybrid exchange-correlation (xc) functional PBE0\cite{PBE0} augmented by a many-body dispersion (MBD) correction,\cite{MBD} using \texttt{tier 2} basis sets and \texttt{tight} settings which resulted in 52 (23) conformers in the low-energy region, \textit{i.e.} within $3\,\mathrm{kcal/mol}$ from the global minimum.

\section{Results and Discussion}
\subsection{\ce{AcPheAla5LysH+}}
For our comparative study, a firm assignment of measured conformer-selective IR spectra to their calculated counterparts is of paramount importance. 
To that end, we first re-assess the peptide \ce{AcPheAla5LysH+} and demonstrate that the applied conformational search technique completely grasps the conformational space energetically close to the global minimum, and that the applied level of theory is capable of reproducing the energetics as well as the vibrational properties of the conformers.
For this we compare our results to previous work on \ce{AcPheAla5LysH+} by Stearns \textit{et al.}\cite{Rizzo2009}, where the 45 lowest-energy structures were selected out of a set of 1,000 force-field minima and subsequently optimized using DFT with a hybrid exchange-correlation (xc) functional.
Even though four structures were successfully assigned to the experimental spectra, the question whether the search was complete and the whether these conformers are located in the global minimum region remained open.
This did, in part, motivate an exhaustive conformational search by Rossi \textit{et al.}~\cite{Rossi2014}, in which $7$ conformers were found within $1\,\mathrm{kcal/mol}$ of the global minimum on the potential-energy surface (PES).
The authors were able to assign the experimentally observed structures to the global minima populated at low temperature by using the hybrid xc-functional PBE0\cite{PBE0}, augmented by a many-body dispersion (MBD) correction,\cite{MBD} and including zero-point energy corrections.
The latter were computed with the generalized-gradient approximation functional PBE~\cite{PBE} and a pair-wise van der Waals correction (vdW),\cite{TS} which proved however unsatisfying for the prediction of vibrational spectra.
It was suggested that using a hybrid xc-functional was necessary, which was a natural assumption since this level of theory was necessary for a correct conformational energy prediction in the first place. 
Furthermore, it was assumed that an anharmonic treatment was needed to yield improved spectra.

\begin{figure}
\includegraphics[width=0.5\textwidth]{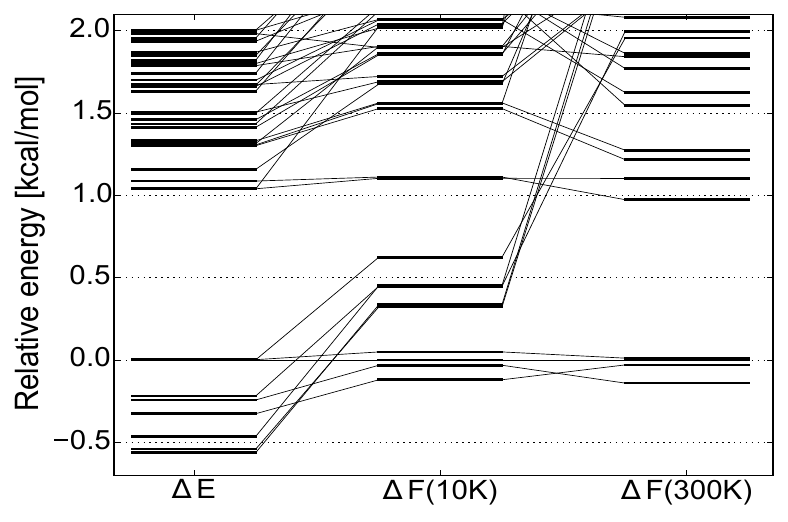}%
\caption{\label{Figure02}Energy hierarchies of conformers of $\mathrm{AcPheAla_5LysH^+}$ at the PBE0+MBD energy $\Delta E$ as well as the Helmholtz free energy $\Delta F$ at $10\,\mathrm{K}$ and $300\,\mathrm{K}$ with harmonic vibrational free energy contributions calculated at the PBE+vdW level.}
\end{figure}

The conformational search strategy has already been laid out in detail in the previous section, including numbers illustrating the exhaustiveness of the search.
The fact that we find two additional conformers within $1\,\mathrm{kcal/mol}$ from the lowest-energy conformer gives us confidence in the conformational search.
The corresponding hierarchy of the relative DFT energy $\Delta E$ on the PES is shown in Figure~\ref{Figure02}. 
Nine conformers were found within $1\,\mathrm{kcal/mol}$ from the global minimum.

\begin{figure}
\includegraphics[width=1.\textwidth]{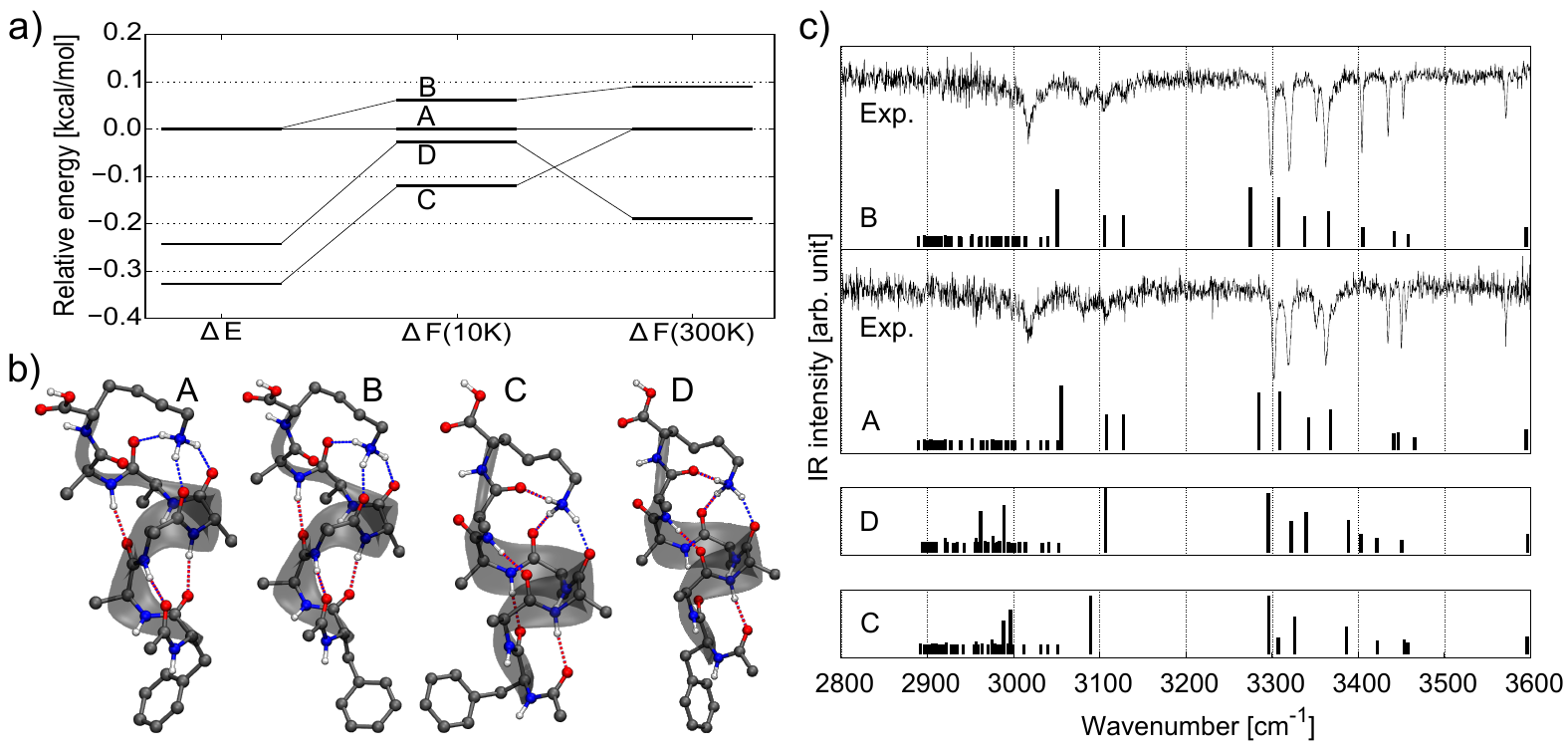}%
\caption{\label{Figure03}
(a)~Relative DFT energies $\Delta E$ as well as relative Helmholtz free energies $\Delta F$ at $10\,\mathrm{K}$ and $300\,\mathrm{K}$ for the lowest-energy conformers of \ce{AcPheAla5LysH+} at the PBE0+MBD level. 
(b)~The four lowest-energy conformers on the $\Delta F(10\,\mathrm{K})$ scale. Hydrogen bonds are indicated with dashed lines. The labeling of the conformers follows Stearns \textit{et al.}\cite{Rizzo2009}.
(c)~Two measured conformer-selective IR spectra (\textit{traces}) are compared to harmonic vibrational calculations (\textit{sticks}). Calculated spectra are uniformly scaled by a factor of $0.948$.}
\end{figure}

Since the experimental measurement takes place on cold ions in the gas phase, the PES merely allows for a rough estimate about the structures populated at low temperatures.
To confidently assign the experimentally observed structures one needs to rely on the Helmholtz free energy $F$ at $10\,\mathrm{K}$, as this is approximately the temperature of the observed ions.
We account for free energy contributions from internal degrees of freedom, consisting of vibrations and rotations, in addition to the DFT energy $E$ on the PES. A detailed formulaic description is provided in the supporting information.
For \ce{AcPheAla5LysH+}, Figure~\ref{Figure02} shows energy hierarchies of the PBE0+MBD energy $\Delta E$ as well as the Helmholtz free energy $\Delta F$ at $10\,\mathrm{K}$ and at $300\,\mathrm{K}$, always relative to conformer \textbf{A} (see Figure~\ref{Figure03}(b)).
At this stage, harmonic vibrational free energy contributions have been calculated at the PBE+vdW level.
While the $\Delta F(10\,\mathrm{K})$ surface should best resemble experimental conditions of gas-phase measurements at $10\,\mathrm{K}$, the free energy hierarchy at $300\,\mathrm{K}$ represents an estimate of the conformers populated at the early stage of the experimental process, where the molecules are electrosprayed into the instrument at room temperature.
Their low free energy at $10\,\mathrm{K}$ and the relatively large gap to alternative structures at $300\,\mathrm{K}$ indicate why the species observed in experiment should be among the four conformers within $0.25\,\mathrm{kcal/mol}$ from the global minimum.
Of course we are aware of the limitation of not taking into account anharmonicity and the possibility of solvation-memory effects (\textit{i.e.} kinetic trapping).

High computational costs prohibited the systematic use of hybrid xc-functionals for the calculation of harmonic vibrations in the previous study by Rossi \textit{et al.}\cite{Rossi2014}
To complete the picture, we repeat the harmonic vibrational free energy calculations at the PBE0+MBD level, confirming the already obtained result. 
Figure~\ref{Figure03}(a) shows the energy hierarchies for $\Delta E$, $\Delta F(10\,\mathrm{K})$, and $\Delta F(300\,\mathrm{K})$ for the four lowest-energy conformers illustrated in Figure~\ref{Figure03}(b). 
Conformers \textbf{A} and \textbf{B} are virtually identical near the C-terminus, but differ near the N-terminus by a tilted Phe side chain.
The difference between conformers \textbf{C} and \textbf{D} is similar. 
All four conformers show helical structure motifs: conformer \textbf{C} possesses one $3_{10}$- and two $\mathrm{\alpha}$-helical turns, conformer \textbf{D} features one $3_{10}$- and one $\mathrm{\alpha}$-helical turn, and conformers \textbf{A} and \textbf{B} each possess two $3_{10}$- and one $\mathrm{\alpha}$ turn.

For this work, the original IR-UV double resonance experiment by Stearns \textit{et al.}\cite{Rizzo2009} has been repeated 
to allow conformer-selective IR spectra to be compared to their theoretical counterparts calculated at the PBE0+MBD level. 
The affiliated UV spectrum including peak assignments to their corresponding conformers is provided in the supporting information.
The conformer-selective IR spectra are shown in Figure~\ref{Figure03}(c).
Conformers \textbf{A} and \textbf{B} could be attributed to their corresponding observed IR spectra. 
While the agreement is very good, the match between experimental and theoretical IR spectra is not perfect. This discrepancy is commonly attributed to two factors: (i)~The effect of a possible incomplete characterization of electron exchange and correlation, despite the use of the hybrid functional PBE0, and (ii)~the treatment of anharmonic vibrations and nuclear quantum effects~\cite{ScalingFactor1}.
Both of these effects are corrected for solely by applying a scale factor to the vibrational frequencies.
The assumption of a \textit{uniform} overestimation of the harmonic vibrational modes with respect to experiment is debatable as they depend on the theoretical method, the used basis set, and the system itself.\cite{ScalingFactor2,ScalingFactor3}
In this work, we focus on the frequency region of $3200\,\mathrm{cm^{-1}}$ to $3500\,\mathrm{cm^{-1}}$ which is sensitive to \ce{N-H $\cdots$ O} hydrogen bonding, where a uniform scaling factor of $0.948$ yields very good agreement.

The exhaustive conformational search presented here for \ce{AcPheAla5LysH+}, and the rigorous treatment of harmonic vibrations at the hybrid xc level allowed for (i)~reproducing the known energy hierarchy and finding additional conformers in the low-energy region and (ii)~calculating well-fitting harmonic IR spectra for the conformers in the low-energy region.
In this way we confirm the conformers predicted by Stearns \textit{et al.}\cite{Rizzo2009} and Rossi \textit{et al.}\cite{Rossi2014} and can rule out any other competing conformers. 
This also shows that calculating computationally costly anharmonic IR spectra is not required in this case. 
Now that we have confirmed the accuracy of our simulation approach, we tackle \ce{AcPheAla6 + Na+}, a more challenging system because of the additional conformational degrees of freedom due to the ``unfixed'' cation. 

\subsection{\ce{AcPheAla6 + Na+}}

\begin{figure}
\includegraphics[width=0.5\textwidth]{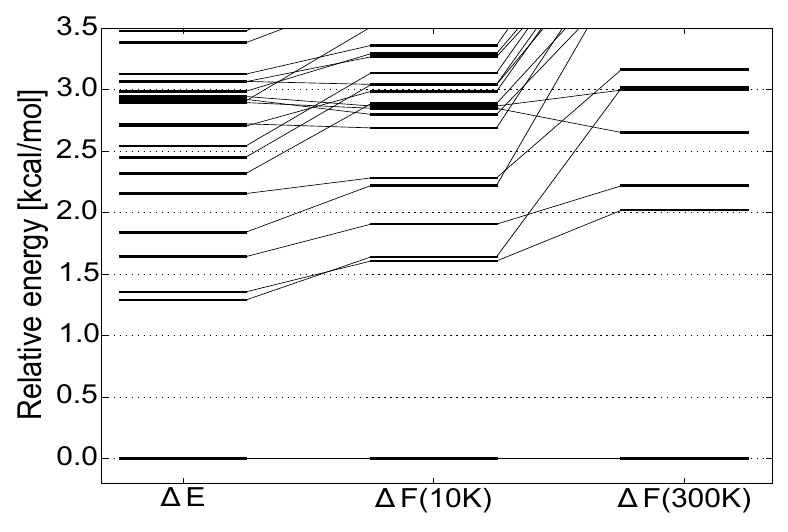}%
\caption{\label{Figure04}Energy hierarchies of conformers of \ce{AcPheAla6 + Na+} at the PBE0+MBD energy $\Delta E$ as well as the relative Helmholtz free energy $\Delta F$ at $10\,\mathrm{K}$ and $300\,\mathrm{K}$ with harmonic vibrational free energy contributions calculated at the PBE+vdW level.}
\end{figure}

\begin{figure}
\includegraphics[width=0.49\textwidth]{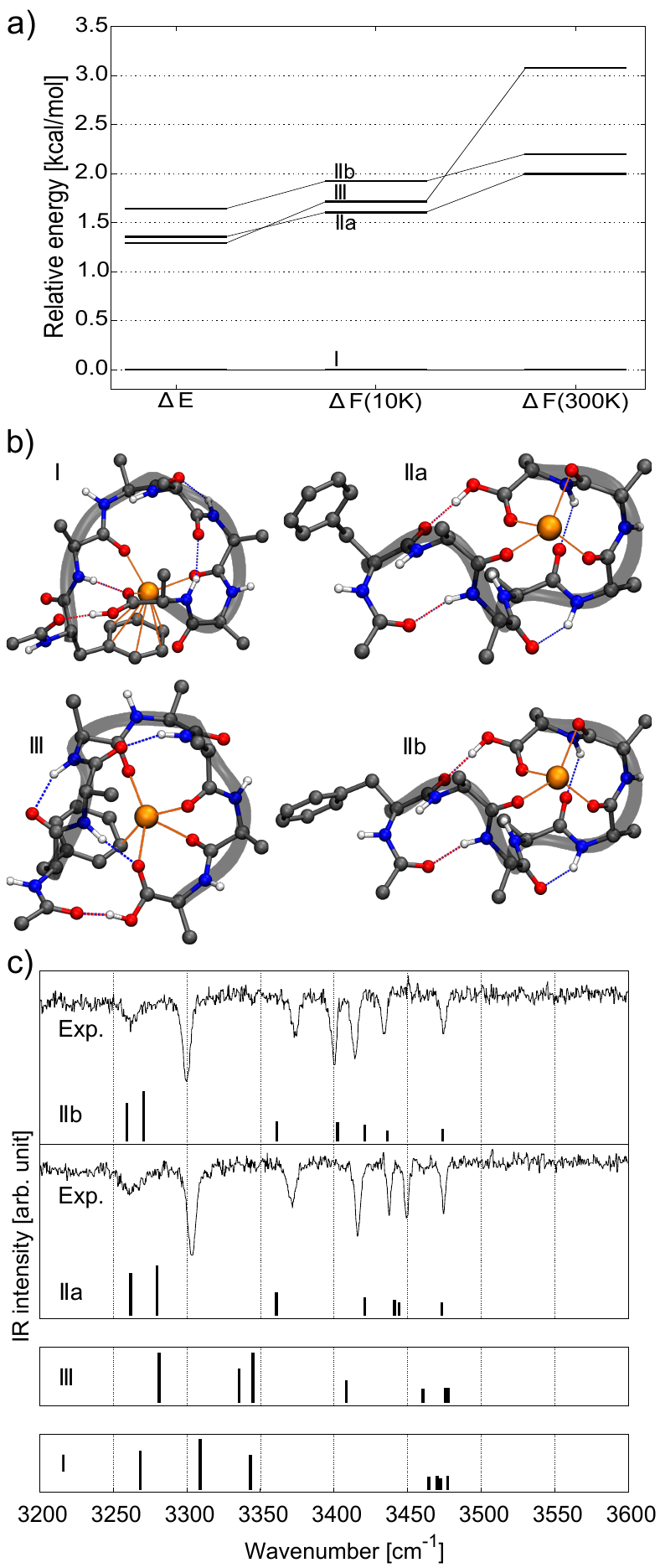}%
\caption{\label{Figure05}
(a)~Relative DFT energies $\Delta E$ as well as relative Helmholtz free energies $\Delta F$ at $10\,\mathrm{K}$ and $300\,\mathrm{K}$ for the lowest-energy conformers of \ce{AcPheAla6 + Na+} at the PBE0+MBD level.
(b)~The four lowest-energy conformers on the $\Delta F(10\,\mathrm{K})$ scale. Hydrogen bonds are indicated with dashed lines. 
(c)~Two measured conformer-selective IR spectra (\textit{traces}) with highest intensity are compared to harmonic vibrational calculations (\textit{sticks}). Calculated spectra were uniformly shifted by a factor $0.948$.}
\end{figure}

Figure~\ref{Figure04} shows the energy hierarchies of the relative PBE0+MBD energies $\Delta E$ as well as the relative Helmholtz free energies $\Delta F$ at $10\,\mathrm{K}$ and $300\,\mathrm{K}$ with harmonic vibrational free energy contributions at the PBE+vdW level that were obtained for \ce{AcPheAla6 + Na+}. 
The four presumably dominant conformers are presented in Figure~\ref{Figure05}(b).
Of the four conformer-selective IR spectra that were recorded, two of them correspond to conformers with particularly high intensity in the UV spectrum (see Figure~S2, supporting information). 
The measured IR spectra of these two conformers, \textbf{\RNum{2}a} and \textbf{\RNum{2}b}, show very good agreement with the IR spectra calculated at the PBE0+MBD, which uses a scale factor of $0.948$.
Both conformers are nearly identical, differing only in the tilt of the \ce{Phe} side chain near the N-terminus.
They are globular with the peptide being ``wrapped around'' the \ce{Na+} cation with four partially negatively charged \ce{C=O} groups pointing towards the positively charged cation, restricting them from forming the hydrogen bonds necessary for helix formation. 
Indeed, no similarities are observed comparing these structures to the helical motifs of \ce{AcPheAla5LysH+}.
The C-terminal fixation of the charge by the \ce{Lys} side-chain seems to be a prerequisite to effectively cap the helix. 
The ``freely movable'' charge prevents helix formation in this system and instead induces a globular motif. 
All conformers found in the low-energy region (\textit{i.e.} within $3\,\mathrm{kcal/mol}$ from the global minimum) show a globular conformation.

An obvious observation is the outstanding global minimum (conformer \textbf{\RNum{1}} in Figure~\ref{Figure05}(b)) that is separated by a $1.6\,\mathrm{kcal/mol}$ gap from the next minimum on the $\Delta F(10\,\mathrm{K})$ scale.
The clear assignment of conformers \textbf{\RNum{2}a} and \textbf{\RNum{2}b} to the two most intense bands in the measured spectra suggests that both conformers may be kinetically trapped.
Moreover, the most stable structure \textbf{\RNum{1}} does not seem to be observed in the experiment -- none of the conformer-selective spectra fit the calculated vibrational signatures (see Figure~\ref{Figure05}(c)).
The structure representing the global minimum is globular and features a cation-$\mathrm{\pi}$ interaction between the \ce{Na+} and the \ce{Phe} side chain. 
If that conformer were present in experiment, one would expect broad features in the UV spectrum due to charge-transfer between \ce{Na+} and the aromatic ring.
However, no such features have been observed.
The reason behind the kinetic trapping of conformers \textbf{\RNum{2}a} and \textbf{\RNum{2}b} has to be sought in the experimental procedure in which the molecules are electrosprayed into the apparatus from a solution at room temperature while the actual measurements are taken on isolated molecules at $10\,\mathrm{K}$.

The energy landscape of the system may differ significantly between the system in solution at room temperature and in the gas phase at $10\,\mathrm{K}$.
This is particularly true for the global minima.
While the global minimum in the gas phase may be energetically favored in comparison to the other structures, kinetic constrains, \textit{i.e.} high energy barriers between minima, may hinder proper folding while transitioning from solution to the gas phase.
Thus, experimentally observed local minima in the gas phase higher in energy are yielded due to their structural bias from aqueous solution at room temperature, resulting in \textit{kinetically trapped} structures unable to transition into the global minimum.

It is obvious from comparing the $\Delta F(10\,\mathrm{K})$ and $\Delta F(300\,\mathrm{K})$ hierarchies (see Figure~\ref{Figure05}(a)) that the temperature difference does not contribute to a possible kinetic trapping effect. 
In fact, the energy gap between the global and the next minimum even increases from $1.6\,\mathrm{kcal/mol}$ at $10\,\mathrm{K}$ to $2.0\,\mathrm{kcal/mol}$ at $300\,\mathrm{K}$. 
Therefore, kinetic trapping must be caused by solvation effects alone.
In order to estimate the magnitude of such an effect, the four lowest-energy conformers presented in Figure~\ref{Figure05} have been geometrically optimized with PBE0+MBD including implicit water by solving the Modified Poisson-Boltzmann (MPB) equation~\cite{MPB1,MPB2} implemented~\cite{MPB3} in \texttt{FHI-aims} (consult the supporting information for computational details).
While in the gas phase conformer \textbf{\RNum{1}} is $1.6\,\mathrm{kcal/mol}$ lower in DFT energy than the next minima (conformers \textbf{\RNum{2}a} and \textbf{\RNum{2}b}), the situation is reversed when including implicit aqueous solution; conformer \textbf{\RNum{1}} is now $0.9\,\mathrm{kcal/mol}$ higher in energy. 
This suggests that they carry a structural bias from aqueous solution, \textit{i.e.} the barriers are sufficiently high to kinetically trap them during the electrospray process.


A similar scenario can be seen for conformer \textbf{\RNum{3}}, which is of comparable energy as conformers \textbf{\RNum{2}a} and \textbf{\RNum{2}b} on the $\Delta F(10\,\mathrm{K})$ scale, but the calculated IR spectrum, presented in Figure~\ref{Figure05}(c), does not match any experimentally observed one. 
Consulting the $\Delta F(300\,\mathrm{K})$ scale (see Figure~\ref{Figure05}(a)) shows that conformer \textbf{\RNum{3}} is $0.9\,\mathrm{kcal/mol}$ higher in energy than conformer \textbf{\RNum{2}b} at room temperature. 
When re-relaxing the structures to the nearest minimum on the potential energy surface at the PBE0+MBD level including implicit aqueous solvation effects as described above, conformer \textbf{\RNum{3}} becomes further energetically penalized -- it is then more than $5.0\,\mathrm{kcal/mol}$ higher in energy compared to the other conformers.


\begin{figure}
\includegraphics[width=1.\textwidth]{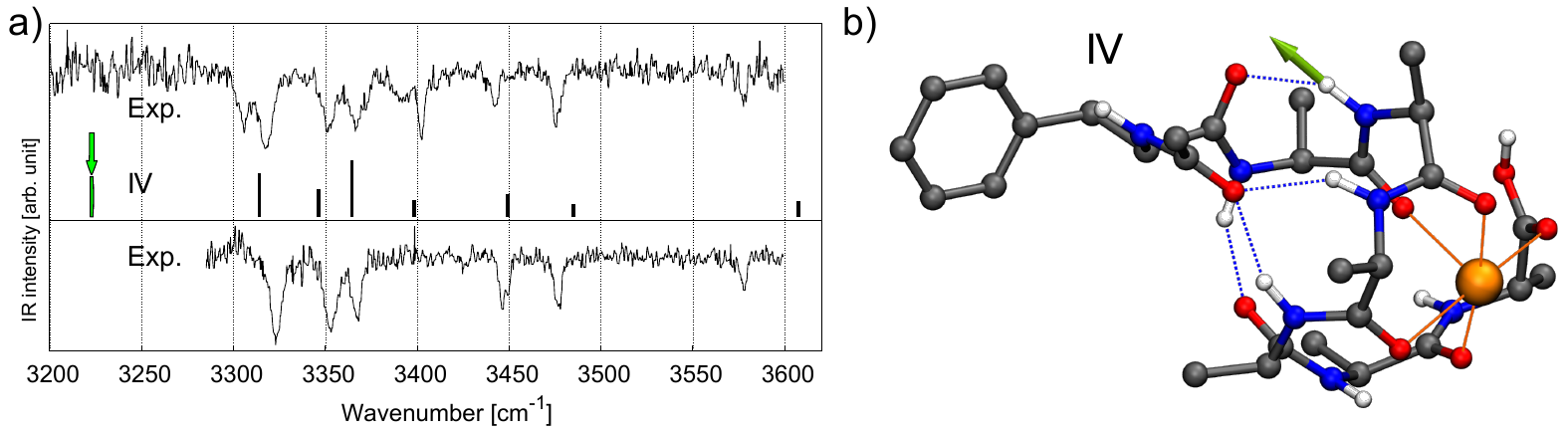}%
\caption{\label{Figure06}(a) For the system of \ce{AcPheAla6 + Na+}, the two measured conformer-selective IR spectra (\textit{traces}) with lowest intensity are compared to vibrational calculations (\textit{sticks}) in harmonic approximation on the PBE0+MBD level for structure \textbf{\RNum{4}}. Calculated spectra have been shifted by applying a uniform scaling factor of $0.948$.
(b)~Structural form of conformer \textbf{\RNum{4}}. Hydrogen bonds are indicated with dashed lines. The highlighted vibrational mode in Figure~(a) is indicated with a green arrow in Figure~(b).}
\end{figure}

There remain two conformers, \textbf{\RNum{4}} and \textbf{\RNum{5}}, for which the UV spectral signatures have lower intensity (see Figure~S2), suggesting that they have smaller populations.
The corresponding IR spectra, shown in Figure~\ref{Figure06}(a), could not be assigned to their calculated counterparts for any structure within $6\,\mathrm{kcal/mol}$ from the global minimum on the $\Delta F(10\,\mathrm{K})$ scale.
Similarly, as for \textbf{\RNum{2}a} and \textbf{\RNum{2}b}, we assume that these conformers are kinetically trapped, which also renders their assignment difficult in assigning them as these conformers might be higher in energy, and thus no energy criterion can be applied for finding them. 
Instead we follow an approach\cite{Voronina2016} where we make use of information from the experiment in order to select from the overall pool of structures for calculation of spectra. 
Candidates were picked if they feature a free carboxylic acid \ce{OH} stretch, since the experimental IR spectra show a peak at $3578\,\mathrm{cm^{-1}}$ (see Figure~\ref{Figure06}(a)). 
Due to the absence of broad features in the UV spectrum, only structures were considered where the \ce{Na+} cation was not in close proximity to the phenyl ring. 
In total, vibrational spectra for 126 conformers have been calculated. 
In addition to that, local refinement on the PBE0+MBD level for all 52 found minima structures within $3\,\mathrm{kcal/mol}$ from the global minimum for the system of \ce{AcPheAla5LysH+} has been laid out after formally replacing \ce{Lys} with \ce{Ala + Na+}, with the sodium cation being placed at the position of the amino group nitrogen. Vibrational spectra for the resulting 28 conformers (after clustering) have been calculated as well.
As explained above, computationally-costly hybrid xc-functionals are required in order to gain enough accuracy. 
Only conformer \textbf{\RNum{4}} (see Figure~\ref{Figure06}(b)), lying $13.6\,\mathrm{kcal/mol}$ higher in energy than the global minimum on the $\Delta F(10\,\mathrm{K})$ scale, could be assigned to one of the less populated conformers.
However, one peak in the simulated vibrational spectrum is blue shifted by $80\,\mathrm{cm^{-1}}$ with respect to the nearest experimental peak, and the corresponding vibrational mode is indicated in Figure~\ref{Figure06}(b) with a green arrow.
Conformer \textbf{\RNum{4}} is a candidate for the kinetically trapped structure only because of the (partially) matching IR spectra. 
Taking into account the large computational effort taken, a more appropriate and computationally affordable technique for finding kinetically trapped conformers would be certainly desirable.

\section{Conclusion}
Our data indicate that the fixed location of the charge at the C-terminus is imperative for helix formation in peptides of this length in isolation, as this stabilizes the structure through a cation-helix dipole interaction.
In the case of the freely-movable sodium cation, the cation-backbone and cation-$\pi$ interactions seem to be stronger, leading to local distortions of peptide structure, preventing helix stabilization.
It is interesting to note the high barriers that seem to be involved in interconverting one structure to another. 
Even though the cation-$\pi$ interaction is energetically favored for the \ce{AcPheAla6 + Na+} in the gas phase, the system remains kinetically trapped in a structural state that is characterized by cation-backbone interactions and that is energetically preferred in polar solvent.

\begin{acknowledgement}
The authors thank the joint Max-Planck-EPFL Center for Molecular Nanoscience and Technology for financial support. 
CB thanks Dr. Mariana Rossi for sharing her knowledge about theoretical vibrational spectroscopy and Prof. Matthias Scheffler for his continuous support.
The experimental work was supported by the EPFL as well as the Swiss National Science Foundation through grant 200020\_165908.
\end{acknowledgement}

\section{Supporting information available:}
\begin{itemize}
  \item A detailed and technical description of the experiment and of the applied simulations methods as well as additional results can be found in the following sections.
  \item Numerical details for all shown energy hierarchies as well as experimental and calculated IR spectra, \textit{i.e.} for Figures~\ref{Figure02}--\ref{Figure06} and Figure~10, as well as corresponding xyz-files of conformers are provided under this link:\\
\begin{footnotesize}
\url{http://w0.rz-berlin.mpg.de/user/baldauf/carsten_pdf/Schneider_2017_Helices_SI.zip}
\end{footnotesize}
\end{itemize}

\clearpage

\section{Supporting Information}

\subsection{Experimental setup in detail}

The experimental setup has been described in detail elsewhere~\cite{Rizzo2010}.
Gas-phase protonated peptides are produced in a continuous fashion by nano-electrospray ionization from a $0.1\,\mathrm{mM}$ solution in 50:50 methanol-water. 
The ions enter the instrument through a metal-coated borosilicate capillary and are focused by an ion funnel. The molecules are pre-trapped in a hexapole in order to generate ion packets and to match the duty cycle of the experiment. 
A quadrupole mass filter selects the $\mathrm{m/z}$ of the ions of interest, which are deflected $90\degree$ using an electrostatic bender, guided though an octopole and deflected $90\degree$ a second time before passing through a set of decelerating lenses and injected in a cold octopole ion trap ($4\,\mathrm{K}$). 
Here, they are cooled down by collisions with cold He gas that is pulsed in before their arrival. 
The He pressure is between $6\cdot10^{-6}$ and $10^{-5}\,\mathrm{mbar}$. 
infrared (IR) and ultraviolett (UV) beams are focused inside the trap and used to spectroscopically interrogate the cold molecules.
The charged fragments produced following UV absorption of the parent ions are extracted from the trap and deflected by a third electrostatic bender and passed through a quadrupole mass filter which selects a particular $\mathrm{m/z}$ ratio before they are detected by a channeltron electron multiplier.
The electronic signature of the ions is recorded monitoring the number of fragments for a particular photofragmentation channel as a function of the UV wavenumber. 
Each conformer present has a characteristic UV signature, so the recorded spectrum is a superimposition of lines coming from all conformations of the parent ion that may be present in the trap. 
Fixing the wavenumber of the UV laser and scanning the wavenumber of an infrared laser pulse that arrives $200\,\mathrm{ns}$ earlier allows for acquiring a vibrational spectrum of whatever conformer is resonant with the UV laser.
When the IR pulse is in resonance with a vibrational transition of the ion, part of the population is removed from the ground state to vibrationally excited states, leading to a decrease in UV induced fragmentation, and as the IR wavenumber is scanned one obtains a conformer-specific vibrational spectrum. 
Performing the same experiment on each line of the electronic spectrum allows for assigning each UV spectral feature to a particular conformer.

\subsection{Conformational search approach in detail}

The applied conformational search algorithm is similar to the one used by Rossi \textit{et al.}\cite{Rossi2014} 
First, a global conformational search is performed on the force field (FF) level using the two empirical fixed point charge models of \texttt{CHARMM22}~\cite{CHARMM22} and \texttt{OPLS-AA}~\cite{OPLS1,OPLS2}, separately.
To that end, a basin-hopping approach~\cite{BasinHopping} was applied using the \texttt{scan} program of the \texttt{TINKER} molecular modeling package~\cite{TINKER1,TINKER2}.
To be detailed, \textit{all} torsional modes were taken into consideration and default search parameters were used, \textit{i.e.} an energy threshold for local minima of $100\,\mathrm{kcal/mol}$ and a convergence criterion for local geometry optimizations of $0.0001\,\mathrm{kcal/mol\cdot\text{\AA}}$.
For the system of \ce{AcPheAla5LysH+} (\ce{AcPheAla6 + Na+}) $603\,280$ ($626\,829$) conformers were found using \texttt{CHARMM22} and $643\,938$ ($635\,120$) conformers were found using \texttt{OPLS-AA}. 
Single-point energy calculations on the PBE+vdW level of density-functional theory (DFT) using \texttt{tier 1} basis sets and \texttt{light} settings have been performed for \textit{all} these FF conformers.
All DFT calculations were done using the all-electron/full-potential electronic structure code package \texttt{FHI-aims}~\cite{FHIaims,FHIaimsRI,FHIaimsELPA}.
For the two FFs individually, the 500 conformers with the lowest FF energy and the 500 conformers with the lowest DFT energy, \textit{i.e.} a grand total of 2000 conformers, have been selected.
The 2000 selected conformers were then geometrically relaxed at the PBE+vdW level using \texttt{tier 1} basis sets and \texttt{light} settings.
Relaxation was accomplished using a trust radius method version of the Broyden-Fletcher-Goldfarb-Shanno (BFGS) optimization algorithm~\cite{BFGS}.
After convergence, a clustering scheme was applied in order to rule out duplicates.
To be precise, root-mean-square deviations (RMSD) of atomic positions between any two conformers were calculated using \texttt{OpenBabel}~\cite{OpenBabel}.
Hierarchical clustering was then achieved by applying the Unweighted Pair Group Method with Arithmetic Mean (UPGMA)~\cite{UPGMA} method implemented in \texttt{Python}'s \texttt{SciPy}~\cite{SciPy} library.
Following that, further relaxation was accomplished at the PBE+vdW level using \texttt{tier 2} basis sets and \texttt{tight} settings.
After clustering, this resulted in 324 (159) conformers for the system of \ce{AcPheAla5LysH+} (\ce{AcPheAla6 + Na+}) in the low-energy region, \textit{i.e.} within $6\,\mathrm{kcal/mol}$ from the global minimum. 
These conformers were then again locally refined on the PBE0+MBD level using \texttt{tier 1} basis sets and \texttt{light} settings. 
After clustering, further geometry relaxation on the PBE0+MBD level using \texttt{tier 2} basis sets and \texttt{tight} settings resulted in 52 (23) conformers in the low-energy region, \textit{i.e.} within $3\,\mathrm{kcal/mol}$ from the global minimum.

\subsection{Conformational search approach comparison}

\begin{figure}
\includegraphics{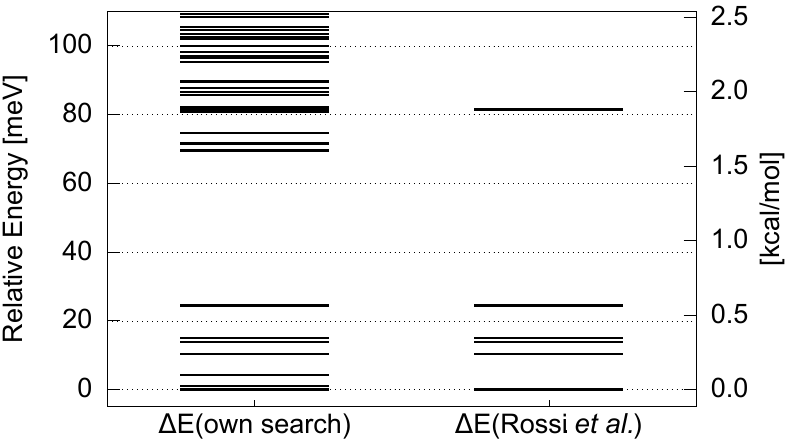}%
\caption{\label{FigureSI02}Comparison of energy hierarchies on the PBE0+MBD level (\texttt{tier 2} basis sets and \texttt{tight} settings) between our own conformational search and the search performed by Rossi \textit{et al.}~\cite{Rossi2014}. Two additional conformers are found in the low-energy region, \textit{i.e.} within $1\,\mathrm{kcal/mol}$ from the global minimum. Conformers with the same energy in both hierarchies correspond to virtually identical structures.}
\end{figure}

For the system of \ce{AcPheAla5LysH+}, nine conformers were found in the low-energy region, \textit{i.e.} within $1\,\mathrm{kcal/mol}$ from the global minimum, two more than Rossi \textit{et al.}~\cite{Rossi2014}. 
Figure~\ref{FigureSI02} compares the two hierarchies on the potential energy surface, \textit{i.e.} on the PBE0+MBD level using \texttt{tier 2} basis sets and \texttt{tight} settings. 
Conformers with the same energy in both hierarchies correspond to virtually identical structures.

\subsection{Helmholtz free energy}

The Helmholtz free energy $F$ per molecule in the gas phase is given by
$$F = E + F_{\mathrm{int}}$$
with E denoting the DFT energy on the potential energy surface (PES) and $F_{\mathrm{int}}$ denoting the free energy contribution due to the internal degrees of freedom, consisting of vibrations and rotations. Assuming harmonic approximation for the intramolecular PES and neglecting any rotational-vibrational coupling, the internal free energy is given by
$$F_{\mathrm{int}} = F_{\mathrm{vib}} + F_{\mathrm{rot}}$$
with
$$F_{\mathrm{vib}} = \sum^{3N-6}_{i}\left[\frac{\hbar\omega_i}{2}+k_{\mathrm{B}}T\ln(1-\exp^{-\hbar\omega_i/k_{\mathrm{B}}T})\right]$$
and
$$F_{\mathrm{rot}} = -k_{\mathrm{B}}T\,\ln\!\left[\pi^{1/2}\left(\frac{2k_{\mathrm{B}}T}{\hbar^2}\right)^{3/2}\sqrt{I_xI_yI_z}\right],$$
where $N$ denotes the number of atoms, $\omega_i$ denotes the vibrational frequency of normal mode~$i$, and $I_x$, $I_y$, and $I_z$ denote the moments of inertia along the three axes. The case of $T=0$ defines the zero-point energy correction where $F_{\mathrm{int}} = \sum^{3N-6}_{i}\frac{\hbar\omega_i}{2}$.

The Helmholtz free energy $F$ is formally given by $F=U-TS$, with $U$, $T$, and $S$ denoting the internal energy, the temperature, and the entropy of the system, respectively.
It is related to the Gibbs free energy $G$ through $G=U-TS+pV=F+pV$, with $p$ and $V$ denoting the pressure and the volume of the system, respectively.
We use Helmholtz free energies in this work because the experiment is essentially done at zero pressure.
Furthermore, throughout this work we are exclusively treating \textit{relative} energies, \textit{i.e.} comparing energy differences between different conformers (usually with respect to the global minimum) of the same system. Hence, the the $pV$ term cancels. In other words, $\Delta G = \Delta F$.

\subsection{Observed UV spectra}

\begin{figure}
\includegraphics{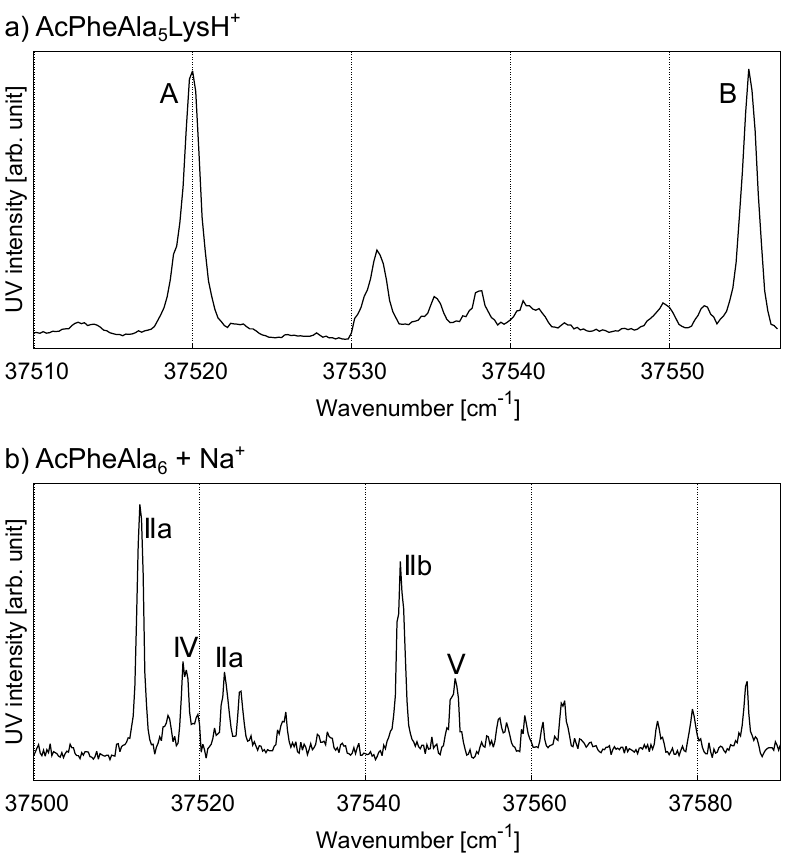}%
\caption{\label{FigureSI03}Measured UV spectra for the systems of (a)~\ce{AcPheAla5LysH+} and (b)~\ce{AcPheAla6 + Na+}. In both cases, peaks have been assigned to their identified conformers shown in Figures~3, 5, and~6.}
\end{figure}

Measured UV spectra for the systems of (a)~\ce{AcPheAla5LysH+} and (b)~\ce{AcPheAla6 + Na+} are presented in Figure~\ref{FigureSI03}. In both cases, peaks have been assigned to their identified conformers shown in Figures.~3, 5, and~6 in the paper.

\subsection{\ce{AcPheAla6 + Na+}: Attempting assignemt of IR spectra of\\[-0.4cm] kinetically trapped conformers \textbf{\RNum{4}} and \textbf{\RNum{5}}}

\begin{figure}
\includegraphics[width=8.0cm]{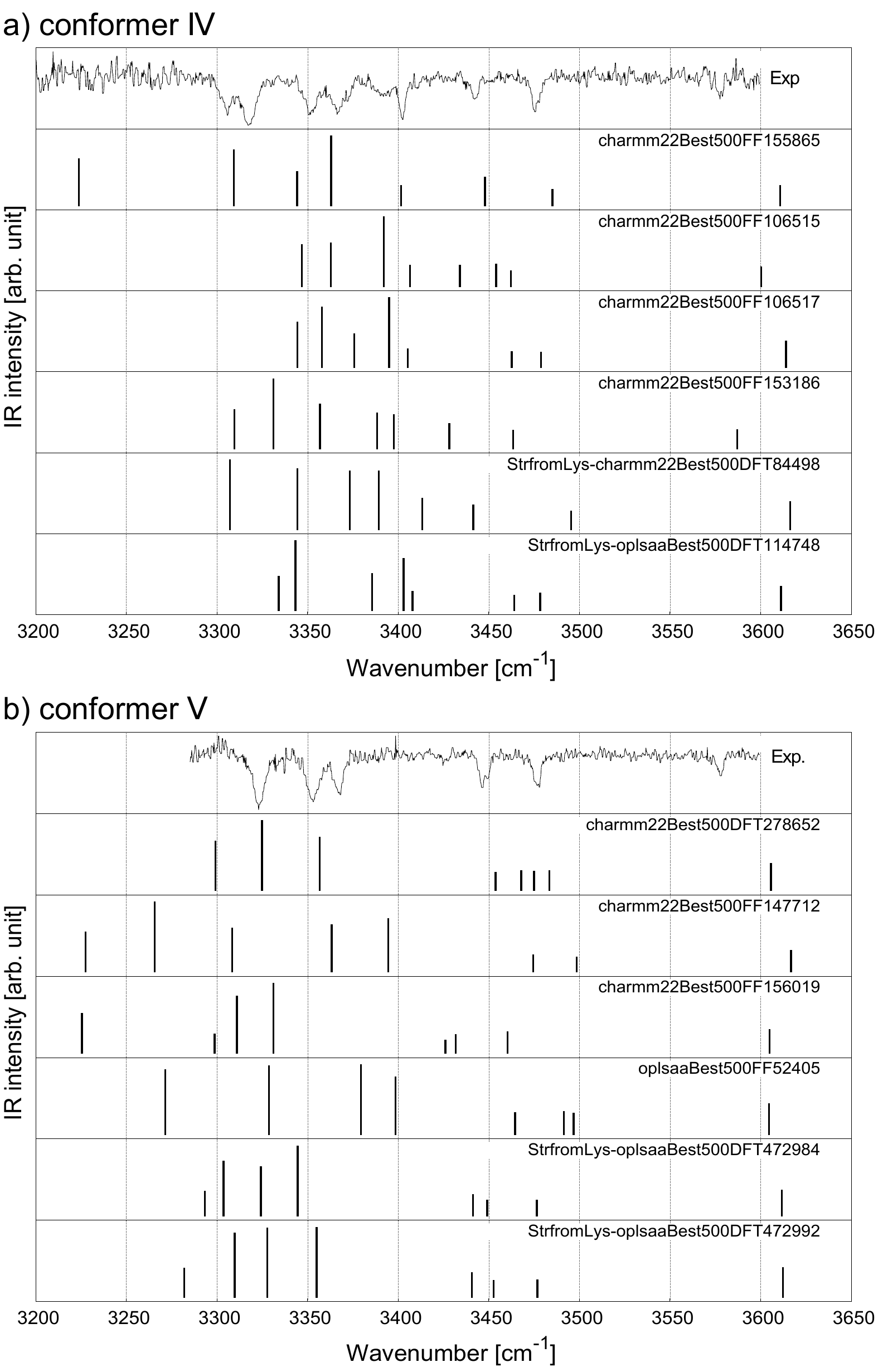}%
\caption{\label{FigureSI05}{Experimentally obtained IR spectra of kinetically trapped conformers (a)~\textbf{\RNum{4}} and (b)~\textbf{\RNum{5}}. In both cases, several calculated IR spectra that share similarities are shown. Conformer \texttt{charmm22Best500FF155865} corresponds to conformer \textbf{\RNum{4}} illustrated in Fioure~6(b) in the paper. \texttt{xyz}-files of the conformers are provided within the \texttt{SI\_data.zip} of this SI.}}
\end{figure}

As stated in the paper, for the system of \ce{AcPheAla6 + Na+} obtained IR spectra of kinetically trapped conformers \textbf{\RNum{4}} and \textbf{\RNum{5}} could not be confidently assigned to their calculated counterparts.
For the sake of completeness, Figure~\ref{FigureSI05} shows the corresponding experimentally obtained IR spectra along with several calculated IR spectra that share similarities.
Calculations have been done on the PBE0+MBD level using \texttt{tier 1} basis sets and \texttt{light} settings.
Conformer \texttt{charmm22Best500FF155865} corresponds to conformer \textbf{\RNum{4}} illustrated in Figure~6(b) in the paper.

\subsection{DFT calculations including implicit solvation effects}

For the system of \ce{AcPheAla6 + Na+} and the four lowest-energy conformers on the $\Delta F(10\,\mathrm{K})$ scale, re-relaxation was applied on the potential energy surface on the PBE0+MBD level (\texttt{tier 2} basis sets, \texttt{tight} settings) including implicit solvation effects by solving the Modified Poisson-Boltzmann (\texttt{MPB}) equation~\cite{MPB1,MPB2} implemented in \texttt{FHI-aims}~\cite{MPB3}. Default parameters have been chosen while explicity setting \texttt{ions\_conc 0} (no ions in the electrolyte). Full relaxation has been achieved for all conformers. Corresponding minima are still fairly similar as the root-mean-square deviation of atomic positions is smaller than $0.5\,\text\AA$ in all cases.

\begin{figure}
\includegraphics{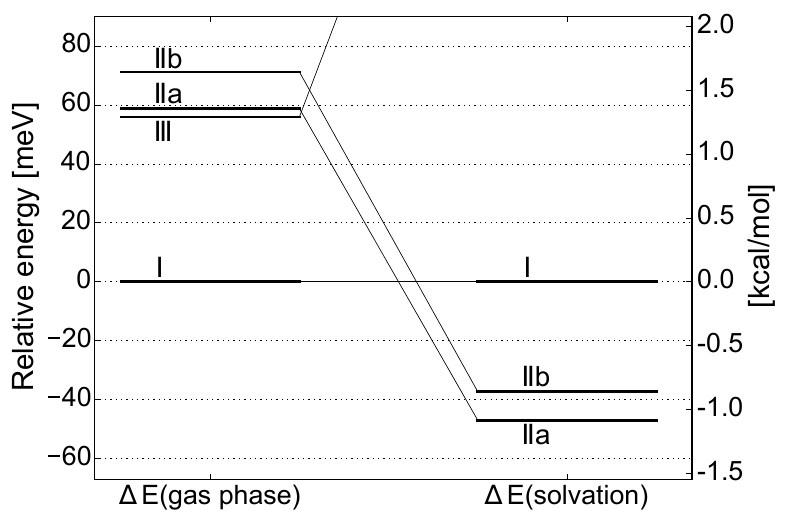}%
\caption{\label{FigureSI01}Comparison of energy hierarchies on the potential energy surface on the PBE0+MBD level (\texttt{tier 2} basis sets and \texttt{tight} settings) between gas-phase calculations and calculations including implicit solvation effects by solving the Modified Poisson-Boltzmann equation (\texttt{MPB}) implemented in \texttt{FHI-aims}. Full relaxation has been achieved for all conformers. Conformers have been labeled as in Figure~5. On the $\Delta E(\mathrm{solvation})$ scale conformer \textbf{\RNum{3}} lies $5.0\,\mathrm{kcal/mol}$ higher in energy than conformer \textbf{\RNum{1}}.}
\end{figure}


\providecommand{\noopsort}[1]{}\providecommand{\singleletter}[1]{#1}%
\providecommand{\latin}[1]{#1}
\providecommand*\mcitethebibliography{\thebibliography}
\csname @ifundefined\endcsname{endmcitethebibliography}
  {\let\endmcitethebibliography\endthebibliography}{}

\end{document}